\input harvmac.tex

\def\half{{1\over 2}}
\def\p{{\partial}}
\def\b{{\beta}}
\def\g{{\gamma}}
\def\d{{\delta}}

\Title{\vbox{\baselineskip12pt
\hbox{IFT-P.001/2009, YITP-SB-09-01, NSF-KITP-09-07}}}
{{\vbox{\centerline{
Regularizing Cubic Open Neveu-Schwarz}
\smallskip
\centerline {String Field Theory}}}}
\smallskip
\bigskip\centerline{Nathan Berkovits${}^{a,c}$ and 
Warren Siegel${}^{b,c}$}
\bigskip\centerline{$a$~~\it Instituto de F\' {\i}sica Te\'orica,}
\centerline{\it S\~ao Paulo State University, S\~ao Paulo, SP 01405-900, Brasil}
\smallskip\centerline{$b$~~\it C.N. Yang Institute for Theoretical Physics,}
\centerline{\it State University of New York, Stony Brook, NY 11794-3840, USA}
\smallskip\centerline{$c$~~\it Kavli Institute for Theoretical Physics,}
\centerline{\it University of California at Santa Barbara, CA 93106-4030, USA}
\bigskip
\vskip .3in

After introducing non-minimal variables, the midpoint insertion
of $Y\bar Y$ in cubic open Neveu-Schwarz string field theory can
be replaced with an operator ${\cal N}_\rho$ depending
on a constant parameter $\rho$. As in cubic open superstring field
theory using the pure spinor formalism, the operator ${\cal N}_\rho$
is invertible and is equal to 1 up to a
BRST-trivial quantity. So unlike the linearized equation of motion
$Y\bar Y QV=0$ which requires truncation of the Hilbert space
in order to imply $QV=0$, 
the linearized  equation ${\cal N}_\rho Q V=0$ directly
implies $Q V=0$. 
\vskip .3in
\Date{January 2009}

\newsec{Introduction}

Open bosonic string field theory has recently been used to study
classical solutions of string theory such as tachyon condensation
that are difficult to analyze using first-quantized approaches.
Some progress has been made in extending these techniques to
open superstring field theory, and there are presently three
versions of open superstring field theory available.

The first version is based on the cubic action\ref\prei{
C.R. Preitschopf, C.B. Thorn and S.A. Yost,
{\it Superstring field theory}, Nucl. Phys. B337 (1990) 363.}\ref\araftwo
{I.Ya. Arafeva and P.B. Medvedev,
{\it Truncation, picture-changing operation and
spacetime supersymmetry in Neveu-Schwarz-Ramond string field theory}, Phys. 
Lett. B202 (1988) 510.}
\eqn\aone{S_1 = \langle Y\bar Y
 ({1\over 2} V Q V + {1\over 3} V * V * V)\rangle
}
where $V$ is the Neveu-Schwarz (NS) string field of zero picture and
$+1$ ghost-number in the small Hilbert space, and $Y\bar Y$ is an
operator of $-2$ picture inserted at the string midpoint.
The second version is based on the Wess-Zumino-Witten-like action\ref
\wzw{N. Berkovits, {\it Super-Poincar\'e invariant superstring field
theory}, Nucl. Phys. B450 (1995) 90, hep-th/9503099.}
\eqn\atwo{S_2 = \langle  (e^{-\phi}Qe^{\phi})(e^{-\phi}\eta e^{\phi})
+\int_0^1 dt \{(e^{-t\phi}Q e^{t\phi})~,~(e^{-t\phi}\eta e^{t\phi})\}
(e^{-t\phi}\p_t e^{t\phi})
\rangle}
where $\phi$ is the NS string field of zero picture and zero ghost-number
in the large Hilbert space.
And the third version is based on the cubic action\ref\pures{N. Berkovits,
{\it Pure spinor formalism as an N=2 topological string}, JHEP 0510
(2005) 089, hep-th/0509120.}
\eqn\athree{S_3 = \langle {\cal N}_\rho 
({1\over 2} \Phi Q \Phi + {1\over 3} \Phi * \Phi * \Phi)\rangle
}
where $\Phi$ is a superstring field of $+1$ ghost-number
in the $GSO(+)$ sector using
the non-minimal pure spinor formalism, and ${\cal N}_\rho = e^{-\rho\{Q,\chi\}}$
is a BRST-invariant regulator inserted at the midpoint which depends
on a constant parameter $\rho$.

Each of these three versions has advantages and disadvantages. The
first version of
\aone\ has the advantage of being cubic,
but has the disadvantage of being singular since the midpoint
insertion $ Y\bar Y$ is not invertible. So the linearized action
$Y\bar Y QV=0$ does not imply $QV=0$ unless one truncates out states
in the kernel of $Y\bar Y$.\ref\araf{I.Ya. Arefeva and P.B. Medvedev,
{\it Anomalies in Witten's field
theory of the NSR string}, Phys. Lett. B212 (1988) 299.}

The second version of \atwo\ has the disadvantage of being non-polynomial,
but has the 
advantage of being non-singular since there are no midpoint insertions.
So the linearized equation of motion is $\eta Q\phi=0$, which implies
$Q V=0$ where $V\equiv \eta\phi$ is in the small Hilbert space.

Finally, the third version of \athree\ has the advantage of being
cubic and non-singular since, unlike the operator $Y\bar Y$ in \aone,
the operator ${\cal N}_\rho$ has no kernel and is invertible.
So the linearized equation of motion ${\cal N}_\rho Q \Phi =0$ implies
$Q\Phi=0$. The disadvantage of the third version is that $\Phi$ includes
the $GSO(+)$ NS and Ramond sectors of the superstring, but does not
include the $GSO(-)$ NS sector and cannot be used to describe tachyon
condensation.

In this paper, we shall propose a fourth version of open superstring
field theory which combines the advantages of the first and third
versions and eliminates their disadvantages. After adding a pair
of non-minimal variables to the NS formalism, it will be possible
to replace the singular operator $Y\bar Y$ of the first version with
a non-singular invertible operator ${\cal N}_\rho$ depending on the non-minimal
variables and on a constant parameter $\rho$.
The action will be 
\eqn\afour{S_4 = \langle {\cal N}_\rho 
({1\over 2} V Q V + {1\over 3} V * V * V)\rangle
}
where $V$ is a NS string field in the zero picture in the
small Hilbert space which is allowed to depend on the non-minimal
variables.

\newsec{Cubic open NS string field theory}

In the absence of operators involving $\d(\g)$, functional integration
over the bosonic $(\b,\g)$ ghost zero modes produces infinities
in the NS tree amplitude. These delta functions can be inserted in
a BRST-invariant manner using the inverse picture-changing operator
\eqn\ipo{Y = c\p \xi e^{-2\phi} \equiv c {{\d(\g)}\over\g} }
where we use the notation
\eqn\vvn{\g= \eta e^\phi,\quad \b = \p\xi e^{-\phi},\quad
\d(\g) = e^{-\phi},\quad {{\d(\g)}\over\g} = \p\xi e^{-2\phi}.}
Note that the OPE's of $e^\phi$ imply that $\d(\g) = \g ({{\d(\g)}\over\g})$.
Since $\g$ has two zero modes on a disk, open string tree amplitudes
require two inverse-picture-changing operators, and it is convenient
to insert the operator $Y\bar Y$ at the midpoint interaction where
$\bar Y = \bar c\bar\p\bar\xi e^{-2\bar\phi}$
is constructed from the antiholomorphic ghosts.

Since $Y\bar Y$
has a non-trivial kernel, e.g. $Y\bar Y c =0$, it is clear
that $Y\bar Y$ is not invertible unless one truncates out states
from the Hilbert space. In order to replace $Y\bar Y$ with
an invertible operator, the first step will be to write
\eqn\invert{Y\bar Y = 4\int dr \int d\bar r \int_{-\infty}^\infty du
\int_{-\infty}^\infty d\bar u ~e^{r c + \bar r \bar c 
+ u\g^2 + \bar u \bar\g^2}}
where $r$ and $\bar r$ are fermionic variables and $u$ and
$\bar u$ are bosonic variables. Note that
\eqn\uex{4\int_{-\infty}^\infty du e^{u\g^2}
\int_{-\infty}^\infty d\bar u e^{\bar u\bar\g^2}
= 4\d(\g^2)\d(\bar\g^2) = {{\d(\g)}\over{|\g|}}
{{\d(\bar\g)}\over{|\bar\g|}} =
 {{\d(\g)}\over{\g}}
{{\d(\bar\g)}\over{\bar\g}} .}

The next step will be to treat $(r,\bar r)$ and $(u,\bar u)$ as
non-minimal worldsheet variables with conjugate momenta $(s,\bar s)$ and
$(v,\bar v)$ by adding them to the worldsheet action 
\eqn\actn{S = S_{RNS} + \int d^2 z (-s\bar\p r - \bar s\p\bar r +
v\bar\p u + \bar v\p \bar u)}
and to the BRST operator
\eqn\brstn{Q = Q_{RNS} + \int dz ~vr + \int d\bar z \bar v\bar r.}
Using the standard quartet argument, the additional terms in $Q$
imply that physical states in the cohomology of $Q$ are independent
of the non-minimal variables. 
It will also be convenient to perform the similarity transformation
\eqn\simmn{Q \to e^{ c(s\p u + \half \p(su)) +
\bar c(\bar s\bar\p \bar u +\half\bar\p(\bar s\bar u))}
 ~Q~
 e^{-c(s\p u + \half \p(su)) -
\bar c(\bar s\bar\p \bar u +\half\bar\p(\bar s\bar u))}}
$$= Q_{RNS} + \int dz [vr + c(\half \p(vu)+v\p u - \half \p(sr) - s\p r)]
$$
$$
+ \int d\bar z [\bar v\bar r + 
\bar c(\half \bar \p(\bar v\bar u)+\bar v\bar \p \bar u 
- \half \bar \p(\bar s\bar r) - \bar s\bar \p \bar r)]$$
so that $(u,v)$ and $(r,s)$ each carry conformal weight $(-\half, {3\over 2})$.

The final step is to define
\eqn\nrho{{\cal N}_\rho = e^{\rho \{Q,\chi\}}
= e^{\rho [ rc+\bar r \bar c + u(\g^2 + {3\over 2} c\p c) + 
\bar u(\bar\g^2 +
{3\over 2} \bar c\bar \p\bar c)]} }
where $\chi = uc+ \bar u \bar c$, $\rho$ is a nonzero constant, and
we have used that $\chi$ has $-{3\over 2}$ conformal weight to compute the
$u c\p c$ term in \nrho. Using a similar computation as in \invert,
it is easy to check that 
\eqn\ntwo{4\int dr\int d\bar r \int_{-\infty}^{\infty}
du \int_{-\infty}^\infty d\bar u{\cal N}_\rho = Y\bar Y}
where the $\rho$ dependence cancels out and the $uc\p c$ term in \nrho\
does not contribute because of the factor of $c\bar c$ in $Y\bar Y$.

As in the regulator used in the non-minimal pure spinor formalism \pures,
on-shell amplitudes cannot depend on $\rho$ since ${\cal N}_\rho=1
+ Q\Omega$ for some $\Omega$. And ${\cal N}_\rho Q V =0$ implies
$QV=0$ since $\nrho$ is easily inverted to
$({\cal N}_\rho)^{-1} = e^{-\rho\{Q,\chi\}}$.

To define a cubic open NS string field theory using ${\cal N}_\rho$,
one needs to allow the string field $V$ to depend both on the 
original NS worldsheet variables $(x^m,\psi^m;b,c,\b,\g)$ and
on the new non-minimal variables $(r,s,u,v)$. Note that 
bosonization is unnecessary since both $V$ and ${\cal N}_\rho$
can be expressed in terms of $(\b,\g)$ and $(\bar\b,\bar\g)$.
The cubic string field theory action is 
\eqn\newa{S = \langle {\cal N}_\rho 
({1\over 2} V Q V + {1\over 3} V * V * V)\rangle
}
where $\langle ... \rangle$ is defined as usual by functional
integration over all the worldsheet variables. 

Since $u$ and $r$
have two zero modes on the disk, their zero mode integration
reproduces \ntwo. So if one writes
$V= V_0 + \widetilde V$ where $V_0$ is independent of the non-minimal
variables, \ntwo\ implies that the terms in $S$ which
are independent of $\widetilde
V$ are the same as in the original cubic action of \aone. 
However, the terms in $S$ which depend on $\widetilde V$ are necessary
for guaranteeing that the linearized equation of motion is equivalent
to $QV=0$. 

To see how $\widetilde V$ contributes to the action, note that rescaling
\eqn\rescale{u\to {u\over \rho},\quad
r\to {r\over \rho},\quad
\bar u\to {\bar u\over \rho},\quad
\bar r\to {\bar r\over \rho},}
$$v\to v\rho,\quad
s\to s\rho,\quad
\bar v\to \bar v\rho,\quad
\bar s\to \bar s\bar\rho,$$
removes the $\rho$ dependence from ${\cal N}_\rho$
and leaves invariant the worldsheet action and BRST operator.
After this rescaling, the string field depends on $\rho$ as
\eqn\scalv{V = \sum_{n={-\infty}}^\infty \rho^{-n} V_n}
where $n$ counts the number of non-minimal fields in $V_n$, i.e.
\eqn\count{[\int dz (uv + rs) + \int d\bar z (\bar u \bar v + \bar r \bar s)~,
~V_n] = n V_n.}
Note that for string fields of finite conformal weight, $n$ is bounded
from below since $(v,s,\bar v,\bar s)$ carry positive conformal weight.

Using \scalv, the action of \newa\ can be expressed as
$S = \sum_{n={-\infty}}^\infty S_n \rho^{-n}$
where 
\eqn\expa{S_n = \langle {\cal N}_{\rho=1}~ (\half\sum_{m=-\infty}^\infty
V_{n-m} Q V_m + {1\over 3}
\sum_{m,p=-\infty}^\infty V_{n-m-p} * V_m * V_p) \rangle.}
As in the cubic action of \athree\ using the pure spinor formalism,
\expa\ involves an infinite chain of auxiliary fields $V_n$
depending on the non-minimal variables. Since the non-minimal variables
include bosons, \scalv\ and \expa\ resemble the construction of
superfields and actions in harmonic superspace. It should be noted
that, unlike the non-minimal variables in the pure spinor formalism
which all carry non-negative conformal weight, the non-minimal variables
$u$ and $r$ carry $-\half$ conformal weight. So $V_n$ for $n>0$
involves states of negative conformal weight which could complicate
computations using level truncation.\foot{
A related difficulty that has recently
been discussed in \ref\kroyter{M. Kroyter, {\it On string fields
and superstring field theory}, arXiv:0905.1170.} is caused by
the nonzero conformal weight of $\chi = u c + \bar u\bar c$.}

Another possible difficulty, which 
is also a difficulty with all the other cubic superstring
field theory actions, is gauge-fixing. Since the midpoint insertion
of ${\cal N}_\rho$ (like the insertion of $Y\bar Y$) involves the $c$
ghost, fixing the $b_0=0$ gauge may be subtle. One could try to implement
alternative gauge choices such as Schnabl gauge or the gauge choices of
\prei, but these also have subtleties. Note that in the cubic
action using the pure spinor formalism, there are difficulties
with gauge-fixing because of the $(\lambda\hat\lambda)$ poles in the
$b$ ghost \pures. The only action that appears to be free of
gauge-fixing difficulties is the Wess-Zumino-Witten-like action
of \atwo\ where one can easily choose the gauge-fixing conditions
$b_0=\xi_0=0$ as in \ref\sen{N. Berkovits, A. Sen and B. Zwiebach,
{\it Tachyon condensation in superstring field theory}, Nucl. Phys. 
B587 (2000) 147, hep-th/0002211.}.

It would be interesting to extend the action of this paper
to include the Ramond sector. Although one can describe the $GSO(+)$
Ramond sector using the pure spinor version of \athree, 
there is no non-singular
action which can covariantly describe both the $GSO(+)$ and $GSO(-)$
Ramond and NS sectors. It is intriguing that the non-minimal
worldsheet fields $(u,v,r,s)$ introduced here have the same statistics and
conformal weights as the non-minimal worldsheet fields $(\widetilde\g,
\widetilde\b,\xi, \mu)$ which were
used in \ref\bigp{N. Berkovits, M. Hatsuda and W. Siegel, 
{\it The big picture}, Nucl. Phys. B371 (1992) 434, hep-th/9108021.}
to allow a more symmetric
treatment of the NS and Ramond sectors.

\vskip 10pt
{\bf Acknowledgements:}
NB and WS would like to thank the KITP conference ``Fundamental Aspects
of Superstring Theory'' where this work was done.
This research was supported in part by the National Science
Foundation under Grant. No. PHY05-51164.
The research of NB 
was also partially supported by
CNPq grant 300256/94-9 and FAPESP grant 04/11426-0, and the
reserch of WS was partially supported by National Science
Foundation Grant. No. PHY-0653342.

\listrefs
\end